\documentclass[preprint,12pt]{elsarticle}

\usepackage{amssymb}

\journal{Nuclear Physics B}

\begin{document}

\def\nuc#1#2{${}^{#1}$#2}
\def\BBz{0$\nu\beta\beta$}
\def\BBt{2$\nu\beta\beta$}
\def\BB{$\beta\beta$}
\def\Tz{$T^{0\nu}_{1/2}$}
\def\Tt{$T^{2\nu}_{1/2}$}
\def\mj{M{\sc ajo\-ra\-na}}
\def\dem{D{\sc e\-mon\-strat\-or}}
\def\mg{M{\sc a}G{\sc e}}
\def\QBB{Q$_{\beta\beta}$}
\def\mBB{$\left < \mbox{m}_{\beta\beta} \right >$}
\def\ge{$^{76}$Ge}

\begin{frontmatter}

\title{Status of the \textsc{Majorana Demonstrator}}

\author[uw]{C. Cuesta}	

\author[lbnl]{N.~Abgrall}		
\author[pnnl]{I.J.~Arnquist}
\author[usc,ornl]{F.T.~Avignone~III}
\author[ITEP]{A.S.~Barabash}	
\author[ornl]{F.E.~Bertrand}
\author[JINR]{V.~Brudanin}
\author[duke,tunl]{M.~Busch}	
\author[uw]{M.~Buuck}
\author[usd]{D.~Byram}
\author[sdsmt]{A.S.~Caldwell}
\author[lbnl]{Y-D.~Chan}
\author[sdsmt]{C.D.~Christofferson}
\author[uw]{J.A.~Detwiler}	
\author[ut]{Yu.~Efremenko}
\author[ou]{H.~Ejiri}
\author[lanl]{S.R.~Elliott}
\author[ornl]{A.~Galindo-Uribarri}	
\author[unc,tunl]{G.K.~Giovanetti}
\author[lanl]{J.~Goett}	
\author[ornl]{M.P.~Green}
\author[uw]{J.~Gruszko}
\author[uw]{I.S.~Guinn}		
\author[usc]{V.E.~Guiseppe}	
\author[unc,tunl]{R.~Henning}
\author[pnnl]{E.W.~Hoppe}
\author[sdsmt]{S.~Howard}
\author[unc,tunl]{M.A.~Howe}
\author[usd]{B.R.~Jasinski}
\author[blhill]{K.J.~Keeter}
\author[ttu]{M.F.~Kidd}	
\author[ITEP]{S.I.~Konovalov}
\author[pnnl]{R.T.~Kouzes}
\author[pnnl]{B.D.~LaFerriere}
\author[uw]{J.~Leon}	
\author[unc,tunl]{J.~MacMullin}
\author[usd]{R.D.~Martin}
\author[unc,tunl]{S.J.~Meijer}	
\author[lbnl]{S.~Mertens}		
\author[pnnl]{J.L.~Orrell}
\author[unc,tunl]{C.~O'Shaughnessy}	
\author[pnnl]{N.R.~Overman}
\author[lbnl]{A.W.P.~Poon}
\author[ornl]{D.C.~Radford}
\author[unc,tunl]{J.~Rager}	
\author[lanl]{K.~Rielage}
\author[uw]{R.G.H.~Robertson}
\author[ut,ornl]{E.~Romero-Romero}
\author[lbnl]{C.~Schmitt}
\author[unc,tunl]{B.~Shanks}	
\author[JINR]{M.~Shirchenko}
\author[usd]{N.~Snyder}	
\author[sdsmt]{A.M.~Suriano}
\author[usc]{D.~Tedeschi}
\author[JINR]{V.~Timkin}
\author[unc,tunl]{J.E.~Trimble}
\author[ornl]{R.L.~Varner}
\author[JINR]{S.~Vasilyev}
\author[lbnl]{K.~Vetter\fnref{ucb}}
\author[unc,tunl]{K.~Vorren}
\author[ornl]{B.R.~White}	
\author[unc,tunl,ornl]{J.F.~Wilkerson}
\author[usc]{C.~Wiseman}		
\author[lanl]{W.~Xu}
\author[JINR]{E.~Yakushev}
\author[ornl]{C.-H.~Yu}
\author[ITEP]{V.~Yumatov}

\address[uw]{Center for Experimental Nuclear Physics and Astrophysics, and Department of Physics, University of Washington, Seattle, WA, USA}
\address[lbnl]{Nuclear Science Division, Lawrence Berkeley National Laboratory, Berkeley, CA, USA}
\address[pnnl]{Pacific Northwest National Laboratory, Richland, WA, USA}
\address[usc]{Department of Physics and Astronomy, University of South Carolina, Columbia, SC, USA}
\address[ornl]{Oak Ridge National Laboratory, Oak Ridge, TN, USA}
\address[ITEP]{Institute for Theoretical and Experimental Physics, Moscow, Russia}
\address[lanl]{Los Alamos National Laboratory, Los Alamos, NM, USA}
\address[JINR]{Joint Institute for Nuclear Research, Dubna, Russia}
\address[duke]{Department of Physics, Duke University, Durham, NC, USA}
\address[tunl]{Triangle Universities Nuclear Laboratory, Durham, NC, USA}
\address[usd]{Department of Physics, University of South Dakota, Vermillion, SD, USA}
\address[sdsmt]{South Dakota School of Mines and Technology, Rapid City, SD, USA}
\address[ut]{Department of Physics and Astronomy, University of Tennessee, Knoxville, TN, USA}
\address[ou]{Research Center for Nuclear Physics and Department of Physics, Osaka University, Ibaraki, Osaka, Japan}
\address[unc]{Department of Physics and Astronomy, University of North Carolina, Chapel Hill, NC, USA}
\address[blhill]{Department of Physics, Black Hills State University, Spearfish, SD, USA}
\address[ttu]{Tennessee Tech University, Cookeville, TN, USA}
\fntext[ucb]{Alternate Address: Department of Nuclear Engineering, University of California, Berkeley, CA, USA}

\begin{abstract}
The~\mj~Collaboration is constructing the~\mj~\dem,~an ultra-low background, 40-kg modular high purity Ge detector array to search for neutrinoless double-beta decay in \ge. In view of the next generation of tonne-scale Ge-based neutrinoless double-beta decay searches that will probe the neutrino mass scale in the inverted-hierarchy region, a major goal of the~\dem~is to demonstrate a path forward to achieving a background rate at or below 1~count/tonne/year in the 4~keV region of interest around the Q-value at 2039~keV. The current status of the~\dem~is discussed, as are plans for its completion.
\end{abstract}


\end{frontmatter}



\section{The \textsc{ Majorana Demonstrator}}
\label{sec1}

The~\mj~\dem~\cite{mjd} is an array of enriched and natural germanium detectors that will search for the neutrinoless double-beta (0$\nu\beta\beta$) decay of $^{76}$Ge. The specific goals of the~\mj~\dem~are:

\begin{itemize}
  \item Demonstrate a path forward to achieving a background rate at or below 1~cnt/(ROI-t-y) in the 4~keV region of interest (ROI) around the 2039~keV~\QBB~of the $^{76}$Ge 0$\nu\beta\beta$-decay.
  \item Show technical and engineering scalability toward a tonne-scale instrument.
  \item Perform searches for other physics beyond the standard model, such as dark matter and axions.
\end{itemize}

\begin {figure}[ht]
\includegraphics[width=0.8\textwidth]{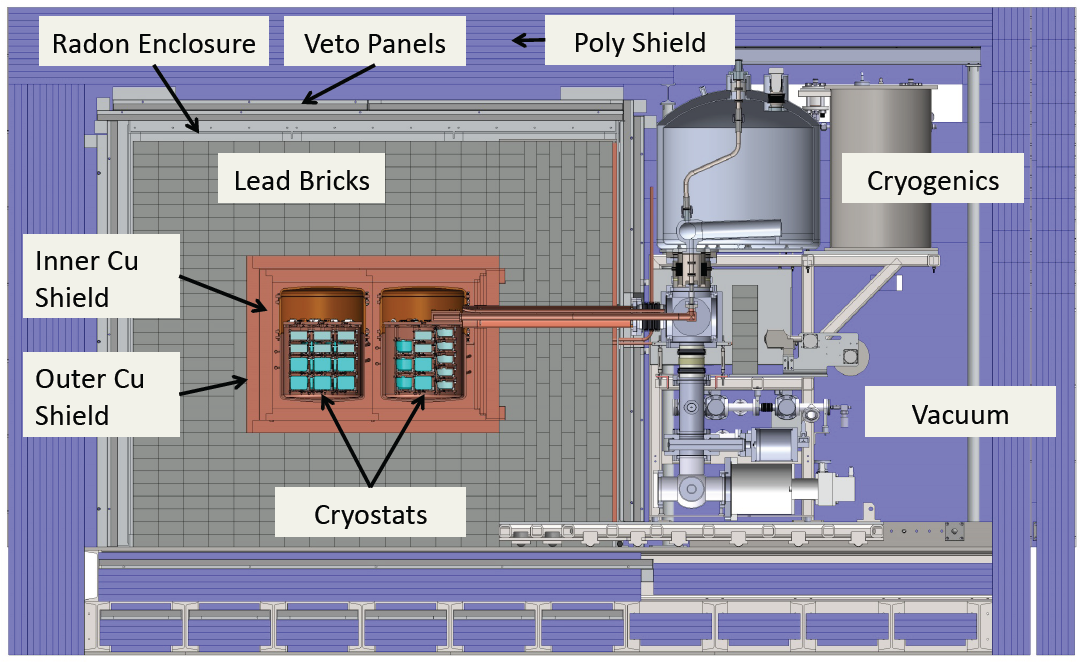}
\centering \caption{\it Schematic drawing of the \textsc{Majorana Demonstrator} shown with both modules installed.}
\label{fig:section}
\end {figure}

\begin {figure}[ht]
\includegraphics[width=0.5\textwidth]{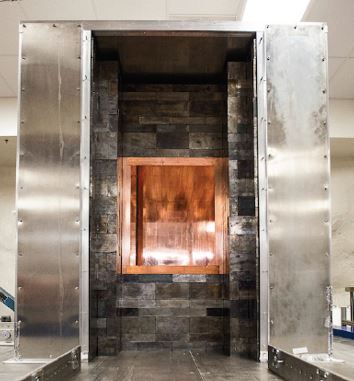}
\centering \caption{\it Low background copper and lead shielding of the \textsc{Majorana Demonstrator} inside the anti-radon enclosure.}
\label{fig:shie}
\end {figure}

The~\mj~\dem~will be composed of 40~kg of HPGe detectors which also act as the source of $^{76}$Ge \BBz-decay. The benefits of HPGe detectors include intrinsically low-background source material, understood enrichment chemistry, excellent energy resolution, and sophisticated event reconstruction. P-type point contact detectors that allow powerful background rejection were chosen after extensive R\&D by the collaboration. The baseline plan calls for 30~kg of the detectors to be built from Ge material enriched to 87\% in isotope 76 and 10~kg fabricated from $^{\rm nat}$Ge (7.8\% $^{76}$Ge). Each detector has a mass of about 0.6-1.0~kg. The main technical challenge is the reduction of environmental ionizing radiation backgrounds by about a factor of 100 below what has been achieved in previous experiments. A modular instrument composed of two cryostats built from ultra-pure electroformed copper has been designed. Each module will host 7 strings of detectors: each detector, with cylindrical shape, is housed in a frame referred to as a detector unit, and up to five detector units are stacked into a string. The modules will be operated in a graded passive shield, which is surrounded by a 4$\pi$ active muon veto. An schematic drawing of the \textsc{Majorana Demonstrator} is shown in Figure~\ref{fig:section} and an image of the low background copper and lead shielding already installed in Figure~\ref{fig:shie}. To mitigate the effect of cosmic rays and prevent cosmogenic activation of detectors and materials, the experiment is being deployed at 4850~ft depth (4260~m.w.e. overburden) at the Sanford Underground Research Facility in Lead, SD.

\section{Status of the~\mj~\dem}
\label{sec2}

The~\mj~\dem~follows a modular implementation to be easily scalable to the next generation experiment. The modular approach will allow the assembly and commissioning of each module independently, providing a fast deployment with minimum interference on already-operational detectors.

\begin {figure}[ht]
\includegraphics[width=0.5\textwidth]{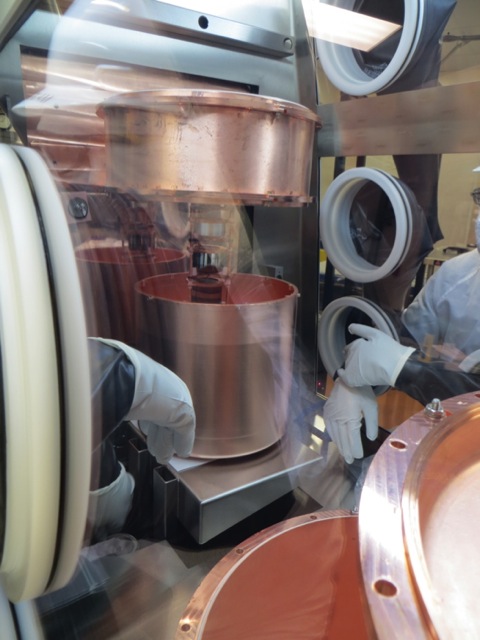}
\centering \caption{\it Prototype Cryostat during the strings installation.}
\label{fig:pm}
\end {figure}

The first step is the Prototype Module, an initial prototype cryostat fabricated from commercially produced copper. At the moment, it is taking data with three strings of detectors produced from natural germanium (see Figure~\ref{fig:pm}) inside the shielding. It has served as a test bench for mechanical designs, fabrication methods, and assembly procedures to be used for the construction of the electroformed-copper Modules~1~\&~2.

\begin {figure}[ht]
\includegraphics[width=0.5\textwidth]{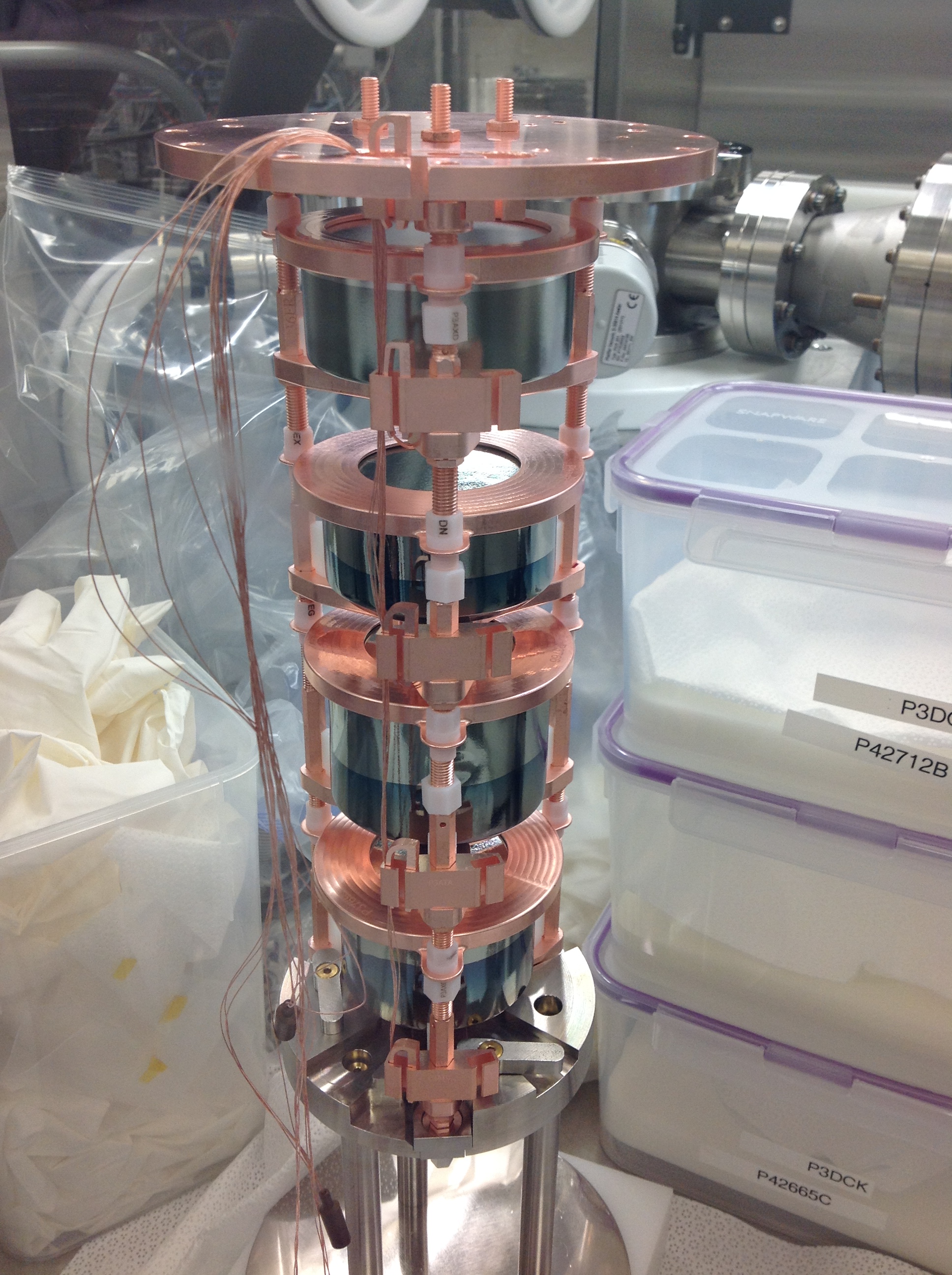}
\centering \caption{\it One string composed by enriched Ge detectors of Module~1.}
\label{fig:string}
\end {figure}

The second step is Module~1, the first module of enriched germanium detectors that is being assembled. The strings (see Figure~\ref{fig:string}) are currently being assembled and characterized in dedicated String Test Cryostats. Module~1 is foreseen to begin the data taking in the beginning of 2015. The last step will be Module~2, composed by enriched and natural Ge detectors. The final assembly of Module~2 will begin once Module~1 is taking data although the parts are already being produced.

\section{Background considerations}
\label{sec3}

The main challenge of the~\mj~\dem~is achieving the background goal: 3~cnts/(ROI-t-y) after analysis cuts that will project to a background level of 1~cnt/(ROI-t-y) in a large scale experiment.  A comprehensive model of expected backgrounds has been constructed~\cite{mjdbkg}, taking into account: Natural radioactivity from detector materials; cosmogenic activation products; backgrounds from the environment or those introduced during detector assembly; muon-induced backgrounds at depth; and the background from atmospheric and other neutrinos. The estimated ROI contributions sum to $<$4.1~cnts/(ROI-t-y) in the~\mj~\dem, and work is in progress to get the final estimate.

One key advantage of HPGe detectors is their inherently excellent energy resolution ($<$0.2\% at $Q_{\beta\beta}$) associated with the low threshold for electron-pair production, leading to a narrow ROI (4~keV for monochromatic 0$\nu\beta\beta$ events). As 0$\nu\beta\beta$-decay produces a mono-energetic peak in the measured spectrum at 2039~keV, improving the resolution reduces the continuum backgrounds in the ROI, allowing for a better identification of lines in the spectrum, and minimizing the contribution from leakage of the irreducible 2$\nu\beta\beta$-decay spectrum into the ROI. The~\mj~\dem~will also make use of event signatures, such as detector coincidences and time correlations, to reject events not attributable to 0$\nu\beta\beta$-decay. Particularly, the P-Type Point Contact design of the~\dem's germanium detectors allow for significant reduction of background through pulse shape analysis.


\section*{Acknowledgments}

This material is based upon work supported by the U.S. Department of Energy, Office of Science, Office of Nuclear Physics. We acknowledge support from the Particle and Nuclear Astrophysics Program of the National Science Foundation. We acknowledge support from the Russian Foundation for Basic Research. We thank our hosts and colleagues at the Sanford Underground Research Facility for their support.

\nocite{*}
\bibliographystyle{elsarticle-num}
\bibliography{NOW14_ccuesta_arxiv}







\end{document}